# Control hubs of complex networks and a polynomial-time identification algorithm

Xizhe Zhang[1], Chunyu Pan[2], Weixiong Zhang[3,4]

1. School of Biomedical Engineering and Informatics, Nanjing Medical University, Nanjing, China

2. School of Computer Science and Engineering, Northeastern University, Shenyang, China

3. Department of Health Technology and Informatics, The Hong Kong Polytechnic University, Hong Kong, China

4. Department of Computer Science and Engineering, Department of Genetics, Washington University in St. Louis, St. Louis, MO, USA

**Abstract**

Unveiling the underlying control principles of complex networks is one of the ultimate goals of network science. We introduce a novel concept, control hub, to reveal a cornerstone of the control structure of a network. The control hubs of a network are the nodes that lie in the middle of a control path in every control scheme of the network. We present a theorem based on graph theory for identifying control hubs without computing all control schemes. We develop an algorithm to identify all control hubs in $O(N^{0.5}L)$ time complexity for a network of $N$ nodes and $L$ links.

**Introduction**

Understanding the control principle of complex networks is of great importance for many applications. Finding important nodes for controlling complex networks can help understand their behaviors and functions and design optimal control schemes.

The close connection between network control theory and graph theory has already been established[1]. Based on the structural controllability theory[1], a network can be controlled, or be driven from an arbitrary state to the desired state in a finite time, if it can be spanned by a cacti structure[2-4]. A key idea underlying the above theory is that to control a directed network, a node can control one of its outgoing neighbors without considering cycles[5]. Such nodes and controlling directions collectively form *control paths*, each of which has a head node at the beginning of the path, also referred to as a *driver node*, a tail node at the end of the path, and middle nodes if any (Figures 1A). All driver nodes and their control paths constitute a *control scheme* of the network.

It is important to note that the control scheme is typically not unique for a complex network. A node may take distinct positions and thus play distinct roles in different control schemes (Figure 1B). To understand the behavior and function of a node, it is required to find and consider all control schemes, which is a #P-hard problem[6], meaning that no polynomial algorithm is known for the problem.

We aim to understand the importance of a node across all control schemes. We introduce a novel concept, control hub, to delineate those nodes that lie in the middle of a control path of every control scheme of a network. Control hubs are the most vulnerable, thus most important, spots of a network because an influence on any of them may render the network uncontrollable by any control scheme. We present a theorem to characterize control hubs and develop a polynomial-time algorithm for identifying all control hubs without computing all of them.

**Structural controllability**

Consider a directed network $G(V, E)$ with a set of nodes $V$ and a set of edges $E$. The states of

the network can be determined by the following linear time-invariant dynamics equation:

$$\frac{dx(t)}{dt} = Ax(t) + Bu(t)$$

where **x**$(t)=(x_1(t),…, x_N(t))^T$ denotes the state of all nodes at time $t$, $A$ is the transpose of the adjacency matrix of $G$, **u**$(t)=(u_1(t),…,u_M(t))^T$ is the set of external control signals, and $B$ is the input matrix specifying where control signals are applied onto the network. The nodes receiving external signals are called *driver nodes*. To effectively control network $G$, it is required to identify a set of driver nodes for $G$. We are particularly interested in the *Minimum Driver nodes Set* (MDS), which offers full control of the network with minimum cost.

It has been shown that maximum matching[7] can be adopted to identify the MDS of a network[8]. For network $G(V, E)$, all unmatched nodes in a maximum matching form an MDS of $G$, which are also the driver nodes of $G$ under the maximum matching. When applied to the driver nodes, external control signals transmit along the matching edges to form control paths (Figure 1A). The set of control paths and the corresponding MDS form a *control scheme* of $G$. A network may possess multiply control schemes because the maximum matching is not unique for most networks (Figure 1B).

A node may play different roles and carry different importance in distinct control schemes. The importance of a node in a control scheme depends on its position in the control path to where it belongs. It may be the head, the tail, or a middle node of the control path. However, the position of a node may change from one control scheme to another, e.g., a head node in one control scheme may become a middle node in another scheme (node 2 in Figure 1B).

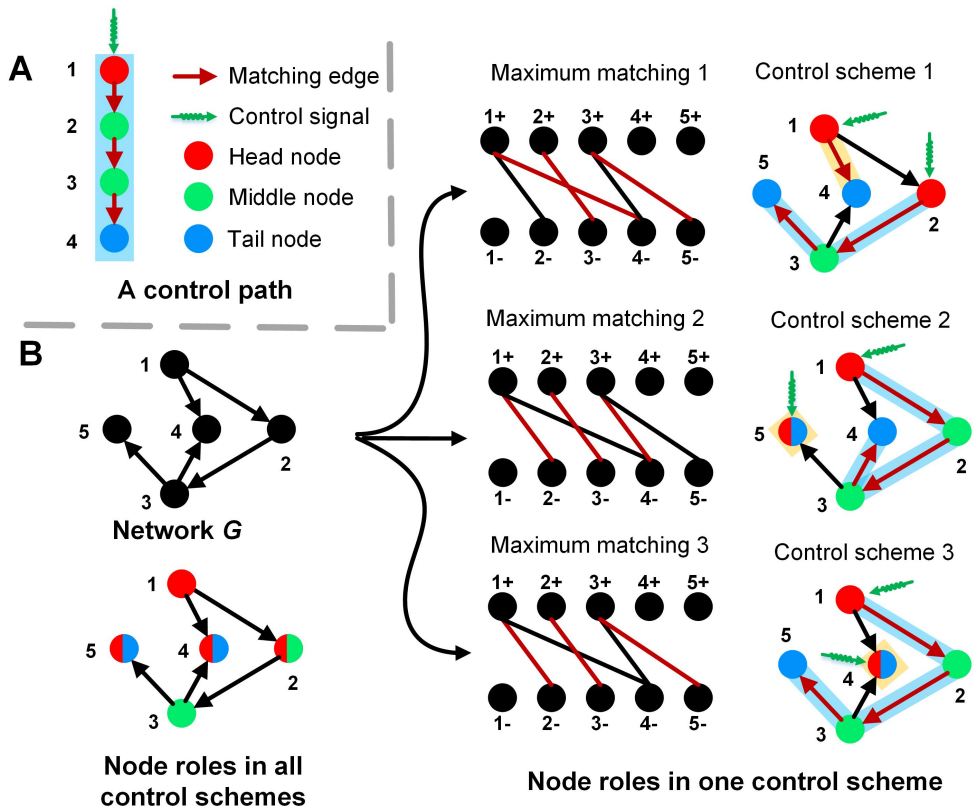

**Figure 1**. The control schemes and control paths of a simple network. **A)** An example of a control path that starts at a head node and ends at a tail node; **B)** A network may have multiple control schemes and a node may take distinct positions in control paths of different control schemes.

**Control hubs**

One type of node is particularly important. Such a node always remains as a middle node of a control path in *all* control schemes and thus is referred to as a *control hub*. An eminent feature of a control hub is that it is essential for controlling the network regardless of which control scheme is applied to the network. A perturbation to any control hub may make the network uncontrollable by any control scheme. Therefore, it is critically important to protect all control hubs to maintain structural controllability.

So, we need to find all control hubs. This seemed to require computing all control schemes, which is #P-hard[6]. We are interested in an efficient algorithm without computing all control schemes.

Consider a bipartite graph $B(V_1, V_2, E)$ with two sets of end nodes $V_1$ and $V_2$ and a set $E$ of edges between $V_1$ and $V_2$. A subset of edges in $E$ is called a *matching M* if no two edges in $M$ have a node in common. A node $v_i$ is said to be *matched* by $M$ if there is an edge of $M$ linked to $v_i$, or otherwise, $v_i$ is unmatched. A *maximum matching* is a matching with the maximum number of edges. A path $P$ is said to be *M-alternating* if the edges of $P$ are alternately in and not in $M$. An *M*-alternating path $P$ that begins and ends at unmatched nodes is called an *M-augmenting path*.

To reiterate, a control hub always resides in the middle of a control path in all control schemes. A node must be a control hub if it is neither a head nor a tail node in any control scheme. Equivalently, a node cannot be a control hub if it can be a head or a tail node in any control scheme. Therefore, to discover all control hubs, we can instead find all the nodes that may be head or tail nodes in any control scheme. We thus have the following properties:

**Property 1**. All head nodes of a network $G$ are the union of the MDSs of all control schemes for $G$.

**Proof**: Since driver nodes receive external control signals, they must be the head nodes of the control paths of a control scheme. Therefore, all the possible head nodes are the union of all the driver nodes of all control schemes of a network.

Based on Property 1, a node is a head node if it is a driver node in some control schemes. Therefore, the head nodes of a network can be easily identified by finding all possible driver nodes for all control schemes, which can be done as we showed previously[9].

**Property 2**[9]. A node is a head node *iff* for an arbitrary maximum matching, it can be reached by an alternating path with an even number of edges that begins at an unmatched node.

Given a network $G$, we say a network $G'$ is the transpose network of $G$ if $G'$ has the same set of nodes but the directions of all edges are reversed.

**Property 3**. The tail nodes of a network $G$ are the head nodes of its transpose network $G'$.

**Proof:** Given network $G(V, E)$, let $B(V_{in}, V_{out}, E)$ be its corresponding undirected bipartite graph. Consider the transpose network $G'(V, E')$ of $G(V, E)$, the only differences between $G$ and $G'$ are the reversed edge directions (Figure 2). Therefore, the undirected bipartite graph of $G'$ is $B'(V_{out}, V_{in}, E)$. Because $B'$ and $B$ have the same set of edges, a maximum matching of $B(V_{in}, V_{out}, E)$ is also a maximum matching of $B'(V_{out}, V_{in}, E)$. Therefore, the control paths of $G$ are the same as that of $G'$ except they have reversed directions and the head nodes of $G$ are the tail nodes of $G'$, and vice versa.

**Theorem 1:** The set of control hubs of a network $G$ is $C=V-H-T$, where $H$ is the set of head nodes and $T$ the set of tail nodes of G.

**Proof**: It follows directly the definition of control hubs and Property 3.

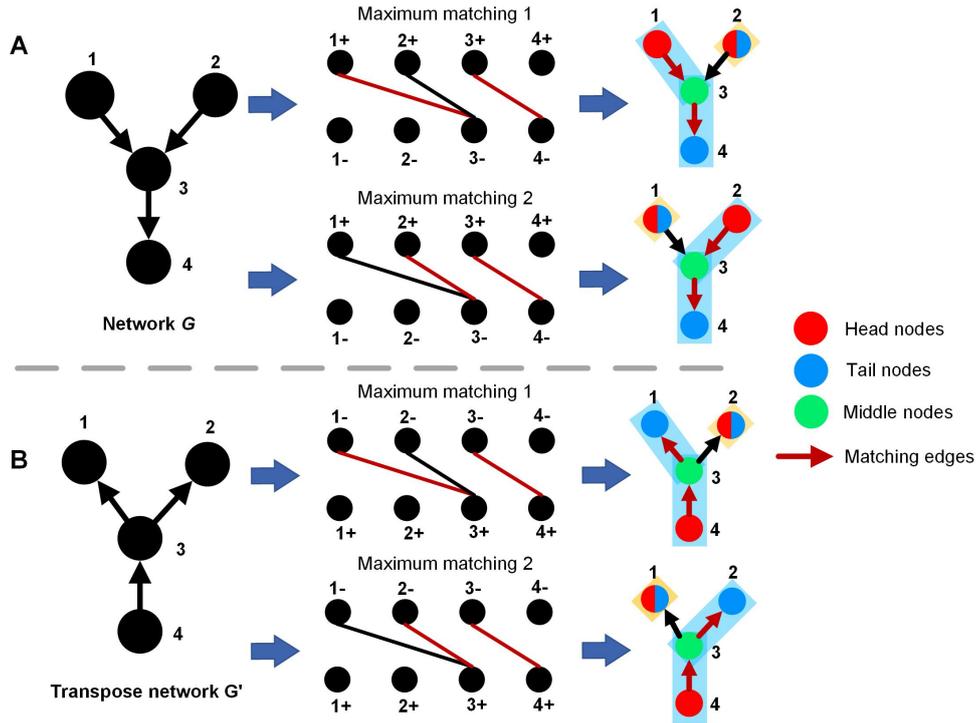

**Figure 2.** Control paths of a simple network and its transpose network. The head nodes of the transpose network of *G* are the tail nodes of *G*. The difference between the two bipartite graphs is that node sets are swapped.

**A polynomial-time algorithm for identifying all control hubs**

Theorem 1 gives rise to a polynomial-time algorithm for finding all control hubs. It entails finding all the possible head nodes of network *G* and its transpose *G'*, which can be done using property 2[9]. Therefore, we have the following algorithm.

**Algorithm: Identification of control hub (G)**

1. Find all possible driver nodes of network *G*(*V*, *E*), denoted as H, by the algorithm in [Zhang, Han, et al. 2017];
2. Find all possible driver nodes of transpose network *G'*(*V, E'*), denote as T;
3. The set of control hubs is the set V-H-T.

Our algorithm for finding all possible driver nodes[9] is listed here for clarity:

**Algorithm: Identification of all possible driver nodes(G)**

1. Construct the bipartite graph $B(V_{out}, V_{in}, E)$ for network *G*; let the initial matching *M* = null;
2. Find all the alternating paths of all unmatched nodes in $V_{in}$, denoted as $AP = \{P_1, P_2…P_n\}$, and let the nodes of *AP* in $V_{in}$ be candidate results;
3. If *AP* contains augmenting paths, expand all augmenting paths and obtain a new matching *M'*; clear all candidate nodes, *M* = *M'*; return to step 2;
4. If *AP* contains no augmenting path, the candidate nodes are all possible driver nodes.

The time complexity of the algorithm for identifying all control hubs is the same as that of the algorithm for identifying all possible driver nodes, which is $O(N^{0.5}L)$ for a network of *N* nodes and *L* links[9].